\documentstyle[psfig]{l-aa}

\begin{document}

\newcommand{\arcm}{\hbox{$^\prime$}}
\newcommand\uarcs{\hskip-0.27em\arcsec\hskip-0.02em}

   \thesaurus{03         % A&A Section: Extragalactic Astronomy:
              (
               12.04.3;  % distance scale
 11.03.4 Virgo Cluster;  % galaxies: clusters: individual
               11.05.1;  % galaxies: elliptical and lenticular
               11.04.1;  % galaxies: distances and redshifts
               11.06.2;  % galaxies: fundamental parameters
               11.16.1.  % galaxies: photometry
                      )} % 

\title{
Distance indicators based on the luminosity-profile shapes of early-type galaxies--a reply
}

%   \subtitle{}

\author{
Christopher Ke-shih Young$^1$ and Malcolm J.\ Currie$^{2,3}$
}

\offprints{C.K.\ Young by e-mailing c.young1@physics.oxford.ac.uk}

\institute{
$^1$Beijing Astronomical Observatory, Chinese Academy of Sciences, Beijing 100080, China\\
$^2$Royal Observatory, Blackford Hill, Edinburgh EH9 3HJ, Scotland, UK\\ 
$^3$Rutherford Appleton Laboratory, Chilton, Didcot, Oxfordshire OX11 0QX, England, UK}

\date{Accepted 1998 January 28}

\maketitle

\markboth{C.K.\ Young \& M.J.\ Currie: Distance indicators based on galaxy profile shapes}{}

\begin{abstract} 
In a recent paper, Binggeli \& Jerjen (1998) question the value of the 
extragalactic distance indicators presented by Young \& Currie (1994 \& 1995) and state 
that they have refuted `the claim that the Virgo dEs [dwarf-elliptical galaxies]...are 
distributed in a prolate structure stretching from 8 to 20 Mpc distance (Young \& 
Currie 1995).' even though no such claim was ever made.

In this paper, we examine Binggeli \& Jerjen's claims that intrinsic scatter rather than
spatial depth must be the main cause of the large scatters observed in the relevant 
scaling relationships for Virgo galaxies. We investigate the accuracy of Binggeli \& 
Jerjen's photometric parameters and find that while their profile curvature and scalelength 
measurements are probably useful, their total magnitude and central surface-brightness 
measurements are not useful for the purpose of investigating scaling laws because they
suffer from serious systematic and random errors.
We also investigate Binggeli \& Jerjen's criticisms of our (1995) analysis. We demonstrate 
that their test for strong mutual dependence between distance estimates based on the two
different scaling laws is invalid because of its prior assumption of negligible 
cluster depth. We further demonstrate that the [relative] distance estimates on which their kinematical 
arguments are based cannot be meaningful, not only because of the seriousness of the 
photometric errors, but also because they are undermined by the prior assumption 
that depth effects can again be neglected.

Interestingly, we also find that Binggeli \& Jerjen's own dataset does itself 
contain evidence for large depth. Using the observed correlation between 
scale-length and profile-curvature, (the only correlation that can be 
investigated meaningfully using their dataset), we find that the frequency 
distribution of residuals with respect to the best fitting curve deviates
significantly from that expected from a uni-modal Gaussian distribution. 
Clearly, if as Binggeli \& Jerjen claim, the very large scatter observed in 
this scaling relationship for Virgo galaxies (which is not observed for Fornax
or Local Group ones) were intrinsic, one would expect a uni-modal Gaussian 
distribution. 

\end{abstract}

\section{Introduction}

Binggeli \& Jerjen (1998) conclude that the shape of a dwarf-elliptical 
galaxy's surface-brightness profile (as quantified by the curvature parameter 
$n$ from S\'{e}rsic's (1968) $r^n$ law) is not a useful distance indicator. 
Their conclusion is based on their finding that the scatter on the relevant 
correlations `can be reduced...never below $\sigma_{rms}\approx0.7$ mag., at 
least for the Virgo cluster.' 

Should the intrinsic scatter on the relevant correlations
be as large as 0.7 mag., the profile-shape indicator would indeed be of only 
limited value, and there would be strong grounds for believing that
Virgo dwarf ellipticals define a single cluster of galaxies of small depth. 
Should however, the intrinsic scatter be about 0.5 mag.\ or lower, profile shape would be 
a valuable indicator of distance and there would be strong grounds for
believing that the Virgo Cluster's depth is significant. Note however, that
the latter interpretation does not require the existence of a `prolate 
structure' as presumed by Binggeli \& Jerjen. There are of course alternative 
models, notably the substructure model we favoured in Young \& Currie (1995)
(hereunder YC95). The central issue in this debate is therefore whether
Binggeli \& Jerjen have demonstrated that the intrinsic scatter in the
scaling relationships is, as they claim, of the order of 0.7$+$ mag., or
whether or not they have at least put forward strong circumstantial evidence in
support of their case.

\section{Limits on the generality of the indicators}

As Binggeli \& Jerjen have mis-interpreted certain aspects of the
luminosity-$n$ ($L$-$n$) and scalelength-$n$ ($R$-$n$) distance indicators of 
Young \& Currie (1994) (hereunder YC94) and YC95, 
it would probably be pertinent to re-emphasize the scope of the 
indicators and how they are related to one another. 

The indicators can be interpreted as follows. Dwarf and intermediate 
elliptical galaxies of the same $n$ are approximately the same physical size
[i.e.\ they have similar $R(n)$]. Only therefore, when such galaxies have 
very similar stellar populations can they be expected to have similar 
central and mean surface brightnesses, whence similar
luminosities\footnote{A very deep sample of galaxies should therefore not 
exhibit a tight correlation between central surface-brightness ($\mu_0$) 
and $n$. It could, however, exhibit a bright-end cut off defined by those 
objects with particularly luminous stellar populations.}. 
It follows that galaxies of different colours can be expected 
to have different stellar populations and therefore cannot be compared
directly using the $L$-$n$ method. The converse is not always true however, as
objects of the same overall colour may have different stellar populations.
The $L$-$n$ relationship appears to be most useful for those dwarf ellipticals 
with colours of $(B-V) \approx 0.7$, because other dwarfs are generally  
bluer (i.e.\ $(B-V) \lse 0.7$) regardless of whether they are higher or 
lower surface-brightness objects (YC95).  

Another important caveat is that if the stellar populations within a galaxy
are not well mixed, the surface-brightness profile shape may deviate 
significantly from that which one might expect on the basis of its size. The 
$n$ based distance indicators should therefore, ideally, not be applied to galaxies with 
internal colour gradients. Although the absence of any colour gradient 
within a galaxy does strictly not imply that its stars are well mixed, in such cases it is 
probably safe to assume that they are. This is because a conspiracy of many
different factors would be required in order to balance the colour gradients
that would otherwise inevitably arise from segregated stellar populations.
With the above in mind, the Local Group early-type dwarf, NGC~205, 
should not be used as an absolute-distance calibrator for $n$-based
scaling laws if those target galaxies possessing colour gradients can be screened out.  
However, it can be used as a calibrator if, as in YC94 and YC95, colour-gradient
information is not available for target galaxies.

Since the distance indicators were first presented, Graham et al.\ (1996)
and Gerbal et al.\ (1997) have found that the correlations on which they are
based probably apply not only to dwarf and intermediate ellipticals, but also
to classical ones, including the brightest cluster ellipticals. Also, Binggeli 
\& Jerjen have also shown that the correlations probably apply to dwarf 
lenticulars as well, while de Jong (1996) has demonstrated that even the bulges 
of spiral galaxies appear to exhibit a continuous range of $n$ values.

It should be remembered however, that colour gradients are more common and often 
much larger in classical ellipticals and lenticulars than in dwarfs. When dealing with 
samples of classical early-type galaxies, it is therefore even more important to 
screen them for objects with colour gradients as such objects cannot be expected to 
conform to the $R$-$n$ relationship [or the $L$-$n$ relationship].

\section{How useful is Binggeli and Jerjen's dataset?}

\begin{figure}
\psfig{figure=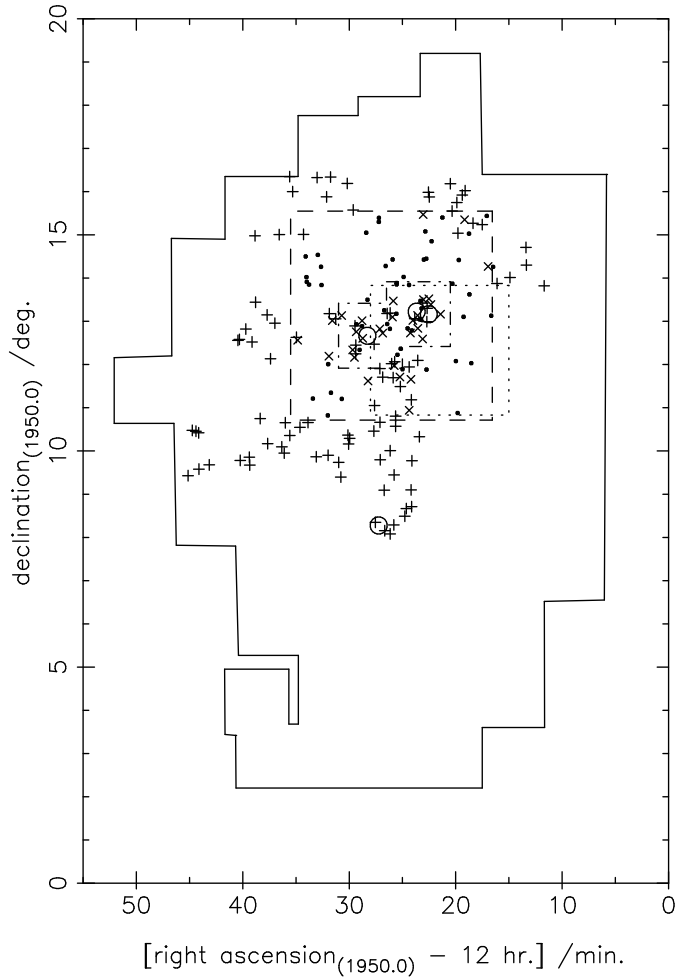,height=13.25cm,angle=-90}
\caption{The distribution of Binggeli \& Jerjen's 128 dwarf galaxies on
the sky: `$\times$' symbols if also YC95 objects or `$+$' symbols if not 
YC95 objects. YC95 objects not in Binggeli \& Jerjen's sample are plotted as 
`.' symbols and the four giant four early-type galaxies: M49, M84, M86 and 
M87, are plotted as `$\circ$' symbols for reference. The largest polygon (solid
line) represents Binggeli et al.'s (1985) {\em Virgo Cluster Catalog\/} survey area
while the largest square (dashed line) represents our (1998) {\em Virgo Photometry 
Catalogue\/} survey area. The areas covered by B\"{o}rngen (1980 \& 1984) and 
Ichikawa et al.\ (1986) are outlined with dotted and dashed-dotted lines 
respectively.}
\label{coverage}
\end{figure}

Binggeli \& Jerjen presented S\'{e}rsic profile parameters for 128 dwarf
elliptical and dwarf lenticular galaxies, which they derived from the
surface photometry of Binggeli \& Cameron (1991 \& 1993). The 
distribution of their objects on the sky is shown in Fig.~\ref{coverage}, 
from which it is evident that only to the south-west of the cluster core
direction (defined collectively by M84, M86 and M87) is 
there detailed coverage.

\begin{figure}
\psfig{figure=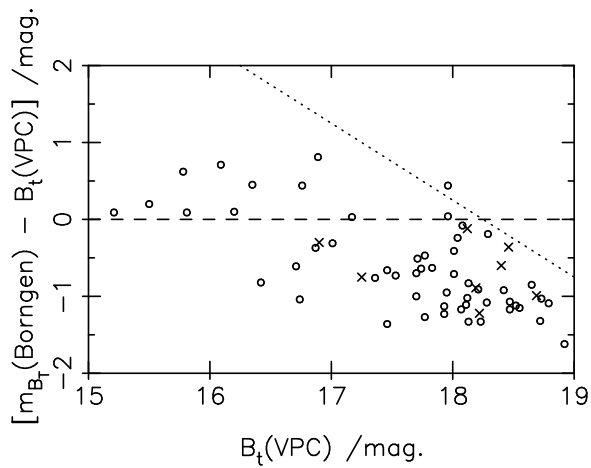,height=7cm,angle=-90}
\caption{A comparison between the $B$-band magnitude scales of B\"{o}rngen 
(1984) and the VPC, based on the 62 galaxies in common between the two samples. 
Binggeli \& Cameron used 13 of B\"{o}rngen's objects as standards for 
calibrating their Plates~18 and 26. Eight of these calibrators were also VPC 
objects, and are depicted as `$\times$' symbols, while the remaining 54 VPC 
objects are shown for reference as `$\circ$' symbols. Note that the large
scale discrepancy must extend to the faint end despite the faint-end limit
to the galaxy sample of $m_{B_T}\sim18.25$ mag.\ (dotted line). This is
because the data points at $17.5<B_{t}(\rm VPC)<18.5$ are concentrated
well below the dotted line (by $\sim1.0$ mag.).}
\label{Bo-VPC}
\end{figure}

\begin{figure}
\psfig{figure=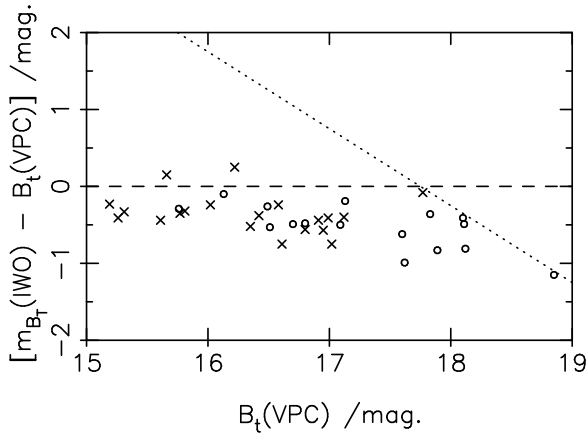,height=7cm,angle=-90}
\caption{A comparison between the magnitude scales of Ichikawa et al.\ (1986)
(IWO)
and the VPC, based on the 36 early-type galaxies in common between the two 
samples. Binggeli \& Cameron's sample of dwarf galaxies had 33 objects 
in common with Ichikawa et al.'s sample, and the mean zero-point discrepancy
based on their objects in common was only 0.04 mag.\ (Binggeli \& Cameron's
magnitudes being brighter). The zero points for their Plates~17,
18 and possibly 26 too, were therefore probably heavily influenced by Ichikawa 
et al.'s magnitude scale.
VPC objects in common with Ichikawa et al.'s sample are
depicted as `$\times$' symbols when also common to Binggeli \& Cameron's 
sample, or otherwise as `$\circ$' symbols. Note that there is not necessarily a
large scale discrepancy at the faint end because Ichikawa's et al.'s galaxy
sample is strongly biased against galaxies fainter than $m_{B_T}\sim$17.75 mag.\
(dotted line).}
\label{IWO-VPC}
\end{figure}

\begin{figure}
\psfig{figure=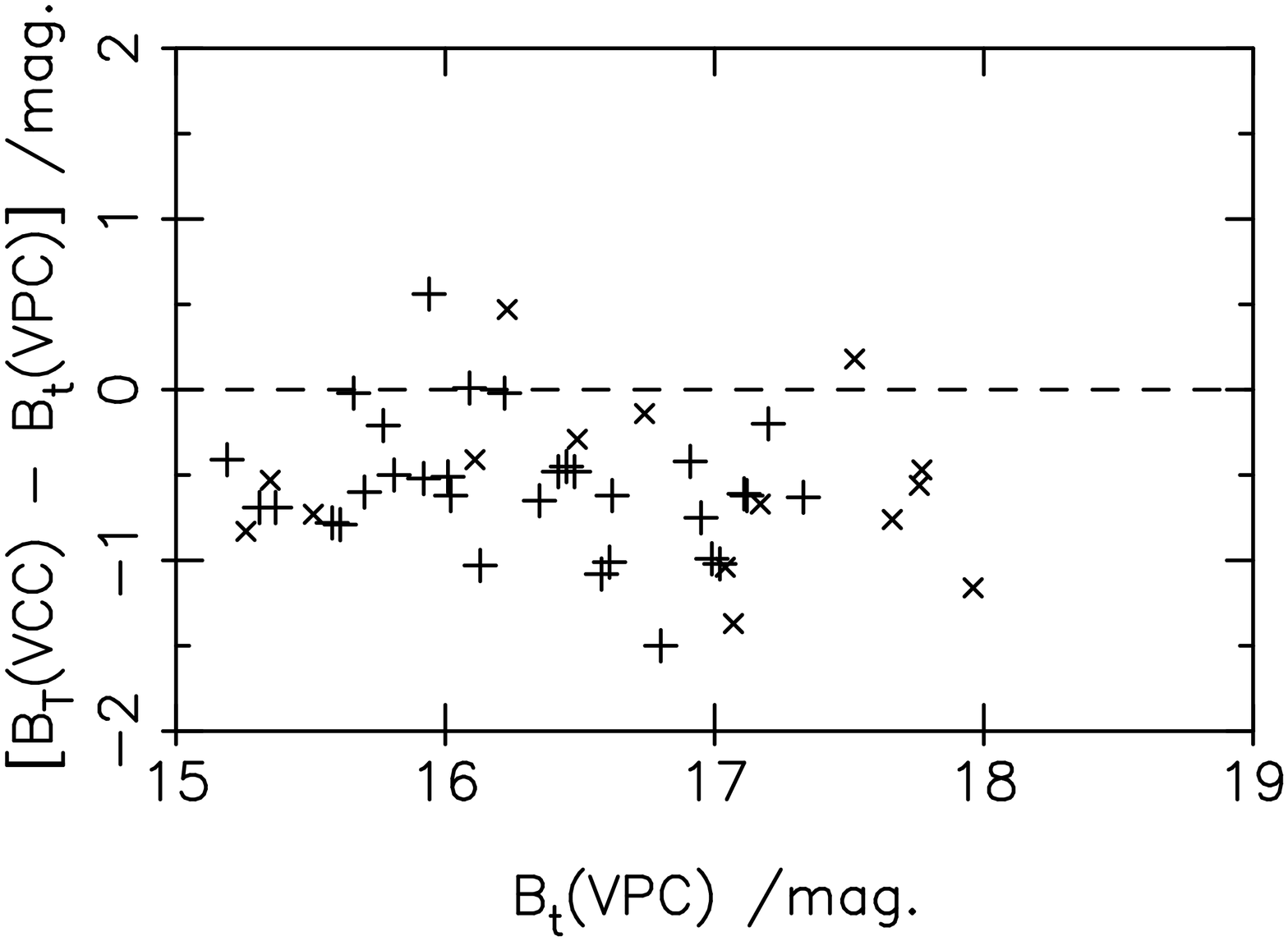,height=7cm,angle=-90}
\caption{A comparison between the magnitude scales of the VCC and the VPC, 
based on the 48 galaxies also common to Binggeli \& Cameron's sample of
dwarfs. Galaxies are depicted as $\times$ symbols unless they are also 
common to Binggeli \& Jerjen's sample, in which case they are depicted
as `$+$' symbols. VCC magnitude values were used by Binggeli \& Cameron
as standards for the determination of their photometric zero-points.
The mean offset is 0.61 mag.\ (VCC values being brighter than VPC ones)
while the scatter is 0.52 mag.}
\label{VCC-VPC}
\end{figure}

Turning now to Binggeli \& Jerjen's total magnitudes, we were surprised to 
discover that the $B_T$ values they quoted were in fact those of Binggeli \& 
Cameron rather than the systemic magnitudes (obtained by integrating
S\'{e}rsic's law to $r=\infty$) that we would have expected. Binggeli \& Cameron's 
photometric zero points were based on the total-magnitude 
scales of de Vaucouleurs \& Pence (1979); B\"orngen (1980 \& 
1984\footnote{Binggeli \& Cameron cited this work as B\"{o}rngen (1983).}) and 
the {\em Virgo Cluster Catalog}, hereunder VCC, of Binggeli et al.\ (1985). 
Ichikawa et al.'s (1986) total-magnitude scale was probably also used 
for the calibration of two or three plates, but Binggeli \& Cameron were 
ambiguous on this point. 

Although, Binggeli \& Cameron were quite modest about the limitations of their 
photometry, Binggeli \& Jerjen allowed for a photometric error of only 
0.2 ~mag.\ in their correlation analyses. There are several reasons why the
real photometric errors must be very much larger than this. These 
reasons are outlined below.

From comparisons with our independently calibrated {\em Virgo Photometry 
Catalogue} (Young \& Currie, 1998)\footnote{This catalogue presents, 
amongst other data for over 1000 galaxies in the direction of the Virgo 
Cluster, $t$-system total magnitudes in the $B_J$ and $B$ bands, as well as 
$U-B{_J}$ and $B_{J}-R_{C}$ colours. 
The $t$ system is described in detail by Young et al.\ (in press), and its
application to the VPC is described by Young \& Currie (1998).
}, hereunder VPC, it is clear that Binggeli \& Cameron's 
adopted magnitude-scale standards do not define a single mutually consistent 
magnitude scale. This point is evident from Fig.s~\ref{Bo-VPC}, \ref{IWO-VPC} 
and \ref{VCC-VPC}, in which serious scale discrepancies between the different 
sources are also noticeable\footnote{
Unfortunately there are no objects in common between the VPC and those
7 standard objects listed in table~1 of Binggeli \& Cameron 
for which de Vaucouleurs \& Pence's total magnitude values were quoted.
This has prevented us from presenting an additional figure here to 
enable comparisons with these extra standards.}.

We are confident that our VPC magnitude scales are 
not to blame for these scale discrepancies, as amongst other reasons, our 
photometry was calibrated with many hundreds of photoelectric 
aperture-photometry as well as simulated aperture-photometry measurements
derived from CCD images.
Furthermore, the agreement between VPC magnitude measurements and those
of Durrell (1997), which were based on deep CCD photometry of relatively faint 
Virgo dwarfs, is better than 0.04 mag. We are also confident that the 
zero point of the VPC's $B$-band total magnitude ($B_t$) scale 
is accurate to several percent (note that it is independent of the 
general transformation we adopted to calibrate our $B_J$ plates with $B$ and 
$V$-band photoelectric photometry measurements). For detailed comparisons
with existing photoelectric photometry for VPC galaxies, see Sect.\ 8 of
Young \& Currie (1998).

Another [albeit related] reason why Binggeli \& Jerjen must have severely 
under-estimated their photometric errors is that Binggeli \& Cameron did not 
calibrate their photometry directly. Instead, they first calibrated their 
extrapolated total-magnitude scale with existing total-magnitude scales. The 
other S\'{e}rsic profile parameters derived by Binggeli \& Jerjen were 
presumably derived on the basis of these magnitude-scale calibrations. The
obvious weakness in this approach is that even if the different sources of
standard objects had been accurately calibrated, there would be systematic  
differences between them on account of the different extrapolation (or in the
cases of B\"{o}rngen's and the VCC datasets, visual total-magnitude estimation) 
procedures.

A further problem is likely to be the scarcity of calibrators in certain 
fields. Binggeli \& Cameron's Plate~1 was for example only calibrated with 
one galaxy.

\begin{figure}
\psfig{figure=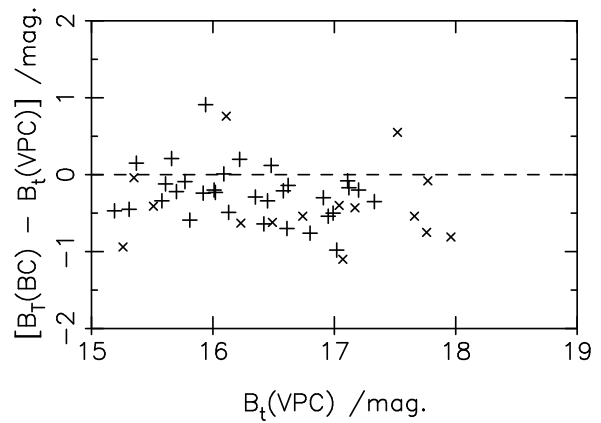,height=7cm,angle=-90}
\caption{A comparison between the magnitude scales of Binggeli \& Cameron
and the VPC, based on the same 48 galaxies shown in Fig.~\ref{VCC-VPC} and
using the same symbols. The mean offset is 0.34 mag.\ (BC values being 
brighter than VPC ones) while the scatter is 0.49 mag.}
\label{BC-VPC}
\end{figure}

If one compares Fig.s~\ref{BC-VPC} with Fig.s~\ref{Bo-VPC}, \ref{IWO-VPC} and
\ref{VCC-VPC}, it is clear that Binggeli \& Cameron's magnitude-scale does not, 
as a whole, bear much resemblance to any of the scales invoked for calibration
purposes. Binggeli \& Jerjen must therefore have severely
under-estimated the errors in both their total magnitude values and 
their central surface brightness, $\mu_0$, measurements.

Young (1994 \& 1997) has already presented some preliminary findings on the 
sizes of and origins of systematic errors in the faint ends of existing 
magnitude scales for Virgo galaxies. A much more detailed paper on this subject,
covering the whole magnitude scale and dealing with the ramifications of
the zero-point and scale errors uncovered, will be presented by Young et 
al.\ (in preparation). 

In spite of the calibration problems described above,
the S\'{e}rsic scalelength, $r_0$, and the shape parameter, 
$n$, should be independent of zero point, so we would expect Binggeli \&
Jerjen's measurements of these quantities to be useful.
Binggeli \& Jerjen noted `quite good' agreement with Durrell's $n$ values, with a 
rms (1$\sigma$) scatter of 0.10. Note that one should not be alarmed by the 
much larger scatter between these authors' log$r_0$ 
values, because this quantity is a strong function of $n$, assuming that 
the $R$-$n$ correlation is genuine (which even Binggeli \& Jerjen
don't question--though they believe that it has a large
intrinsic scatter).

In summary then, Binggeli \& Jerjen's galaxy sample is not a complete sample
of galaxies down to a well-defined total magnitude limit. Also, its coverage 
of the Virgo Cluster direction is very patchy. However, it does contain a
large number of dwarf galaxies and is therefore useful on the basis of 
its size. Unfortunately the photometric zero points adopted for 
different plates are not mutually consistent, thereby rendering the 
$B_T$ and $\mu_0$ values of little use. However, this should not affect 
the $n$ or $r_0$ values, which are probably more accurate than the YC95 
values because they are based on higher-resolution photometry.

\section{Binggeli and Jerjen's correlation analyses} 

Binggeli \& Jerjen investigated the following four correlations: 
$B_T$ versus $\log(n)$, $B_T$ versus $0.712\mu_{0}-3.385\log(n)$, 
$B_T$ versus $\mu_0$, and $\log(r_{0})$ versus $\log(n)$ for Virgo galaxies. 
They observed rms scatters 
in these correlations of 0.92, 0.73, 0.76 and 0.85 mag.\ respectively, and
asserted that: `A scatter of 0.7 mag.\ is what one can already get 
from the relation between the mean effective surface brightness $<\mu>_{\rm eff}$
and total magnitude'.

As is evident from Fig.~\ref{BC-VPC} there is a significant and not necessarily linear
scale error in their magnitude scale for galaxies that lie within the VPC 
survey area (corresponding to their Plates~17, 18 and 26 but with two objects 
on their Plate~4). The sense of this error is such that the luminosities of 
their fainter galaxies were over-estimated with respect to their
brighter objects. In the case of the outlying fields their scale errors are 
almost certainly even larger as the only calibrators 
used were VCC galaxies with total-magnitude values taken from either the VCC or
de Vaucouleurs \& Pence (1979). As mentioned in Sect.~3, there are very large
systematic errors in both of these sources of magnitudes. 
In fact the preliminary work of Young (1994 \& 1997) finds that these
sources over-estimate luminosities by about 0.7 mag.\ at the faint end.

As already demonstrated in Sect.~3, Binggeli \& Jerjen's photometry 
was based on differentially zero-pointed plates
(i.e.\ objects on each of the 13 different plates received different absolute
calibrations). Furthermore, their $B_T$ values were not systemic ones 
(i.e.\ obtained by integrating S\'{e}rsic's function through 360$^\circ$ to 
$r=\infty$), but those of Binggeli \& Cameron (1993), which were obtained
using a different extrapolation procedure and including the nuclear
light contribution when present. The effects of both of these limitations 
in their reduction procedures would be to increase the observed scatter in 
the $B_T$ versus $\log(n)$,
the $B_T$ versus $0.712\mu_{0}-3.385\log(n)$ 
and $B_T$ versus $\mu_0$ correlations. 
Of these three correlations, the first would be affected the most.
This is because the $\mu_0$ term in the other two 
correlations can to a certain extent compensate for the errors
in the $B_T$ values adopted (even if neither the measured $\mu_0$ nor the
measured $B_T$ bear much resemblance to the actual values). Also, the
$B_T$ versus $\log(n)$ correlation is the one most susceptible to increased 
scatter when, as by Binggeli \& Jerjen, applied indiscriminately to objects of 
different stellar populations in the absence of galaxy-colour information.

We therefore find that Binggeli \& Jerjen's dataset is useful only for
investigating the $\log(r_{0})$ versus $\log(n)$ correlation, assuming of 
course that Binggeli \& Cameron's (1993) background subtraction procedures 
were adequate. We are therefore confronted with an observed
scatter of 0.85 mag.\ in a scaling relationship based on a sample of 128 Virgo
galaxies. Clearly, even if the
measurement errors in the parameters $r_0$ and $n$ introduced a 
random component as high as 0.30 mag., we are still left with a scatter
of 0.80 mag.\ to explain. Binggeli \& Jerjen attribute this remaining
component mainly to intrinsic scatter, while we would attribute a large part 
of it to spatial depth.

\section{Dependence of $L$-$n$ and $R$-$n$ distances}

Binggeli \& Jerjen make a big issue of the mutual dependence between the
residuals in magnitude space with respect to their $B_T$ versus $n$ correlation and the
residuals in angular-distance space with respect to their $r_0$ versus $n$ correlation.
They plot these residuals in their fig.~9. While
they are correct in pointing out that there must be some dependence
between the two sets of residuals, whether this dependence is significant
enough to affect our previous findings is another matter.

\begin{figure}
\psfig{figure=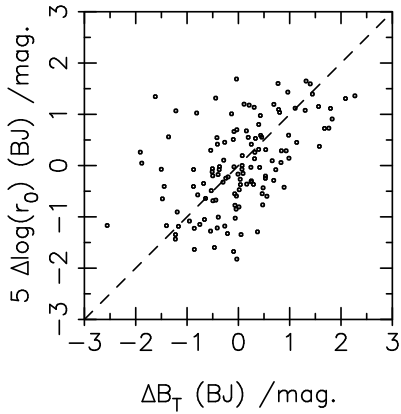,height=7cm,angle=-90}
\caption{Binggeli \& Jerjen cited the {\em strength\/} of this correlation they found using their own
dataset as evidence that the analysis of YC95 was flawed. Their reasoning was that this correlation
must be the product of dependence between distance estimates derived by different methods, rather
than due to genuine depth in the spatial distribution of Virgo galaxies.}
\label{res-BJ}
\end{figure}

We have re-plotted their fig.~9 here as Fig.~\ref{res-BJ}, this time using equal axis scales.
They claim that in the absence of any dependence between the residuals, Fig.~\ref{res-BJ} 
should be devoid of any correlation.
However, their test for dependence is fatally flawed because it is based on the prior 
assumption of negligible depth--as illustrated by the following example.

\begin{figure}
\psfig{figure=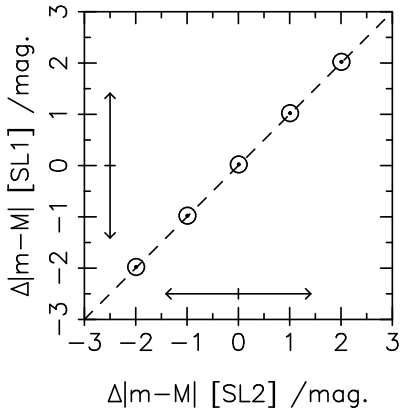,height=7cm,angle=-90}
\caption{This plot is analogous to Fig.~6, but invokes hypothetical galaxy data 
($\odot$ symbols) and two hypothetical distance indicators that are based on 
mutually independent scaling laws, denoted SL1 and SL2. Both indicators are capable 
of yielding precise distance measurements. 
For each indicator, the lengths of the arrows represent the value of the rms scatter 
in the distance residuals with respect to the mean distance obtained for the galaxy sample.
According to Binggeli \& Jerjen, the rms scatters with respect to the equality line 
should both be 2 mag.\ when in fact they are zero!
}
\label{res-hypo}
\end{figure} 

Imagine that we have five galaxies, which collectively constitute a complete
sample of galaxies devoid of any Malmquist bias. The nearest galaxy is at $(m-M)=28$
while the farthest is at $(m-M)=32$, and the spatial separation between
each object is $\Delta(m-M)=1$. The mean distance modulus of these five galaxies 
[in log(distance) space] is therefore $(m-M)=30$. 
Now, let us imagine that we have two perfect distance indicators based on
two completely independent scaling laws which we shall denote
SL1 and SL2. Both indicators can measure the distances of these objects precisely
because both methods are perfect.
If we were now to construct a diagram analogous to Fig.~\ref{res-BJ}, we 
would end up with a plot like Fig.~\ref{res-hypo}. The rms scatter in the
residuals with respect to SL1 would be identical to that with respect to SL2, and
both of these quantities would be equal to $\sqrt{2}$ mag.
Now, according the Binggeli \& Jerjen, for two such `independent but equivalent' measurements,
we would expect the scatter with respect to the equality line on 
Fig.~\ref{res-hypo}
to be 2 mag. However, because the distance indicators are perfect, the
actual scatter with respect to the equality line is {\em zero\/} [regardless of which
axis it is measured parallel to]. The reason for this is that while the two different
measurements for an individual galaxy are `equivalent'; the 
measurements for different objects are not, simply because each object is at a different 
distance.

In spite of the above, we accept that in sect.~4 of YC95, we did indeed 
under-estimate our formal internal distance errors, because there must be some
[non-distance related] dependence between the distance estimates based on the 
different scaling laws. However, even if our formal internal errors [including
both intrinsic scatter and
photometric errors] were as high as, say, 0.6 mag.\ [cf.\ 0.47 mag.\ as quoted
in YC95], that would still leave room for a cluster depth of 0.54 mag., as the
observed scatter per relationship in YC95 was 0.81 mag. Note that a cluster 
with a depth of 0.54 mag.\ would be half as deep as it is distant, with a 
further one-third of its objects lying even further out from its centroid.

We should also like to emphasize that our `independent information on the
intrinsic scatter' was not `seized...by applying both the $n$-$M$ and $n$-$\log r$ 
relation at the same time'. The independent information
was in fact, the much smaller scatter found in our samples 
of Fornax and Local Group galaxies. As long as our Fornax and Local-Group samples are
representative and as long as they contain galaxies structurally similar to their 
counterparts in Virgo, the depth interpretation still holds.

\section{Virgo and Fornax dwarfs: a dichotomy?}

Binggeli \& Jerjen state that: `If the intrinsic dispersion of the $n$-$M$
or the $n$-log$r_0$ relation is much smaller for Fornax dwarfs than for Virgo
dwarfs as it appears (which, however, might be caused by the incompleteness of 
YC's Fornax sample) we are in need of an explanation for this difference'.

In response to their criticisms that we excluded three suspected 
non-cluster members when investigating the scatter in our $R$-$n$ relationship,
we have re-measured the scatter in our
$R$-$n$ correlation without excluding any outlier. For a polynomial of the form
$R = an^{-3} + bn^{-2} + cn^{-1} + d$, which has the advantage over equation~1
in YC95 of being monotonic, the scatter in $R$ based on all 26 of the 
objects listed in table~2 of YC95 is [in terms of magnitudes]: 0.55 mag.
Allowing for a conventional Fornax-Cluster depth of 0.15 mag., but not
making any allowance for possible foreground or background objects, places an
upper limit on the intrinsic scatter of 0.53 mag.
This is very much lower than the scatter found in the same relationship for
Virgo galaxies (see Sect.~4). If, as Binggeli \& Jerjen maintain, the much larger 
scatter observed for Virgo galaxies were intrinsic, we would therefore indeed
be in need of an explanation as to why these Fornax dwarf galaxies 
differ so radically from their counterparts in Virgo. 

As Binggeli \& Jerjen concede, King (1966) profiles do not fit
Virgo dwarf-elliptical galaxy profiles well. This suggests that 
tidal truncation is not a significant contributor to the luminosity profile 
shapes at the radii of interest. We therefore consider it unlikely that tidal
effects could offer the explanation. Furthermore, on the basis of the colour information 
presented by Caldwell \& Bothun (1987) and YC95, it is
clear that most of the brighter Fornax and Virgo dwarf ellipticals have very similar colours, 
suggesting that they may well have very similar stellar populations and histories. We therefore remain to be convinced of
Binggeli \& Jerjen's suggestion that there is probably a dichotomy between Virgo and Fornax dwarfs.

\section{Effective surface brightness versus magnitude}

This relationship was cited by Binggeli \& Jerjen as being of comparable value
to the profile-shape parameter, $n$, as a distance indicator. We do not deny 
that it is a reasonably useful relationship. In fact, it is related to the 
$L$-$n$ and $R$-$n$ relationships, and probably a direct consequence of them.
However, it can be expected to be significantly harder 
than the $L$-$n$ and $R$-$n$ relationships to measure accurately, because it
invokes the effective surface-brightness parameter, which is a tertiary
parameter (unlike $n$ and $r_0$ which are primary parameters and total 
magnitude which is a secondary one). 

In order to measure effective surface brightness accurately,
a model profile must first be fitted, then the profile must be extrapolated 
to obtain a total-light estimate and then the profile model must be integrated 
to the half-light radius. Clearly, an extra stage is involved. We
therefore cannot accept Binggeli \& Jerjen's assertion that no profile 
modeling is required in the measurement of either total magnitude or effective
surface brightness. Young (1997) and Young et al.\
(in press) have already demonstrated that for dwarf galaxies in particular,
total magnitude values (and therefore effective parameters too)
are critically dependent on the profile model adopted.

\section{Cosmic expansion and cluster kinematics}

Binggeli \& Jerjen expected to find a `well-defined velocity-distance
relation' based on their `fairly large' subsample of 43 objects with
known velocities, if there were significant depth in the spatial 
distribution of Virgo dwarfs. They cited the lack of such a relationship
based on their $L$-$n$ and $L$-$\mu_0$ `pseudo-distance' estimates, as
evidence against the depth interpretation. 

\begin{figure}
\psfig{figure=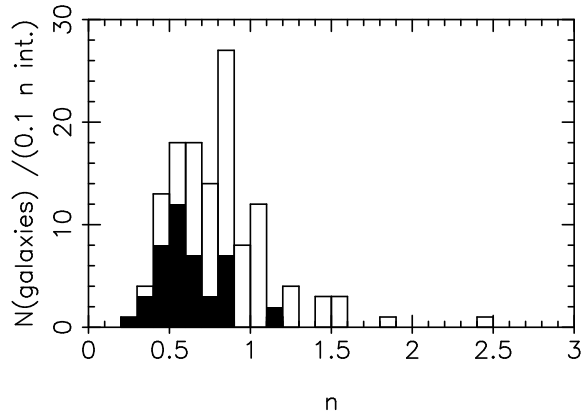,height=7cm,angle=-90}
\caption{A histogram of the number of galaxies in Binggeli \& Jerjen's 
sample of 128 dwarfs per unit profile curvature, $n$, interval. The
shaded regions represent that subsample of 43 objects with published radial 
velocities.}
\label{Malmquist-n}
\end{figure} 

The most fundamental problem with their argument is that
in order to generate relative distances based on each of two different scaling relationships,
they have already assumed negligible depth when they estimate these relationships directly from the residuals
with respect to the best-fitting curves to their data. The crucial point here is that should there be 
significant depth, the mean distance of their high-$n$ objects must be lower than the mean distance
of their low-$n$ objects, due to Malmquist bias. The relative distance scales they construct for each relationship should
therefore not be based on a best fit to data for Virgo galaxies, but on a curve defined by a best fit to
data from either a sample of objects known to be at similar distances (e.g.\ Fornax-Cluster galaxies) 
and/or a sample of objects whose distances are known (e.g.\ Local-Group galaxies).
In Fig.~\ref{Malmquist-n} the differential frequency of galaxies should 
increase monotonically with increasing $n$
if their galaxy sample were unbiased. As this is not what is observed, we can conclude that their
galaxy sample suffers from bias against high $n$ objects.
Should there be depth in the spatial distribution of their sample galaxies,
the mean distance of their lower $n$ objects must be greater than the
mean distance of their higher $n$ objects. 
Their residuals can, at best, therefore only yield meaningful relative distances for 
galaxies within very small ranges in $n$ for which the degree of the
Malmquist bias can be assumed to be constant.

Furthermore, as already demonstrated in Sect.~3, the 
`pseudo-distances' derived by Binggeli \& Jerjen for their fig.s~10 and 11, must indeed be highly
inaccurate on account of the errors in their photometry.
Also, they apply the $B_T$ versus $n$ and $B_T$ versus $\mu_0$
relationships to all early-type dwarfs indiscriminately, in the absence of
e.g.\ colour information. 

The existence of so many outliers (two of which are objects with negative 
radial velocities) on Binggeli \& Jerjen's fig.s~10 \& 11 is therefore not 
surprising. However, as previously suggested in YC95, there may well be 
significant line-of-sight substructure in the spatial distribution of Virgo 
galaxies, complicating the kinematics of the galaxy populations present.
Significant spatial depth therefore need not necessarily imply the `quiet' 
velocity-distance relationship presumed by Binggeli \& Jerjen.

We have been interested in the kinematics of the dwarf-galaxy populations
in Virgo for a number of years now, and a major programme to measure large 
numbers of redshifts for early-type dwarf-candidates is already well underway.
The Virgo galaxies targeted by Drinkwater et al.\ (1996) with the
multi-fibre spectrograph on the United Kingdom Schmidt Telescope (UKST),
were, on account of the observing constraints at the time, generally of 
high surface brightness. Consequently, most of them were found to be in 
the background. However, 8 objects were confirmed to be dwarf or intermediate 
early-type galaxies.
In 1997, early-type objects of low surface-brightness were targeted by
Drinkwater et al.\ (in preparation), again using the UKST, and 67 velocity 
measurements were obtained.
Further measurements made in 1997 with a different telescope and
and future ones (from two separate telescope-time allocations in
1998) will be presented in subsequent papers. 
A detailed investigation into the Virgo-dwarf velocity field will then be 
based on these new data. 

\section{Intrinsic scatter or spatial depth effects?}

\begin{figure}
\psfig{figure=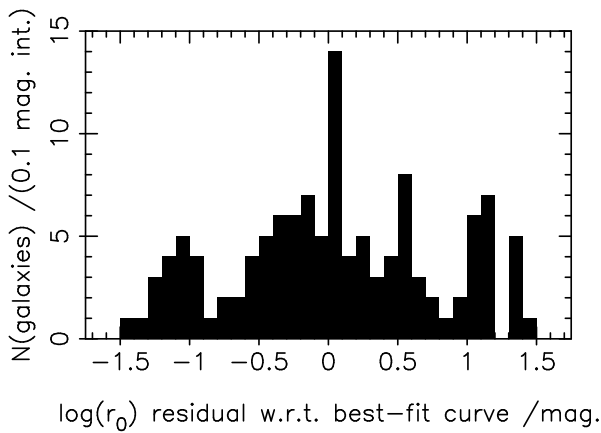,height=7cm,angle=-90}
\caption{The frequency of scale-length residuals with respect to Binggeli \& Jerjen's best fit 
curve to their data as a function of the residuals in magnitude space. 
Note that the abscissa scale cannot be interpreted as a [negative] relative distance scale because no
correction has been made for the differential Malmquist bias effects discussed 
in Sect.~8.}
\label{BF-Rn-res}
\end{figure} 

Even without any new datum, we still have one potentially decisive test that 
might be able to help us decide whether the large scatter observed in Binggeli
\& Jerjen's $\log(r_{0})$--$log(n)$ relationship is due primarily to intrinsic
scatter or depth effects. This test involves looking for departures from 
uni-modality and/or normality in the differential frequency distribution
of scale-length residuals with respect to the best-fit curve for the data 
(not in this case a curve defined by galaxies from an external galaxy sample).
Should the distribution not be consistent with a uni-modal Gaussian [measured 
in log(distance) space to be rigorous], we can say that the scatter is not 
consistent with the intrinsic scatter origin hypothesis. 

In Fig.~\ref{BF-Rn-res}, we find some evidence for a tri or quad-modal 
distribution. Such a distribution could arise if early-type dwarfs exist 
in three or four discrete size ranges within the same cluster. However, not 
only does this seem most unlikely, but it would also be very hard to reconcile
such a scenario with the theoretical work on the subject (Hjorth \& Masden 
1995, Gerbal et al.\ 1997, Prugniel \& Simien 1997 \& Ciotti \& Lanzoni 1997).
It is perhaps more likely that any multi-modality present is due to
line-of-sight substructure in the galaxy sample's spatial distribution, 
albeit smoothed considerably due to the relative distance scales being different 
for different ranges in $n$.

We have performed several non-parametric statistical tests as described by Lucey
et al.\ (1986) and references therein. Comparing the Fig.~\ref{BF-Rn-res} data
against 100,000 Monte-Carlo realisations drawn from a Gaussian distribution
having the standard deviation of the residuals, yielded probabilities
that the Fig.~\ref{BF-Rn-res} distribution could arise by chance.
For the Kolmogorov-Smirnov (Lillefors) test and Geary's a-test (mean absolute
deviation/standard deviation), the probabilities were 0.255 and 0.222
respectively; which were suggestive but not by any means conclusive. 
However, for the skewness and kurtosis tests and the u-test (data range/standard
deviation) the probabilities were 0.041, 0.011 and 0.005 respectively.
We take these results as significant evidence against Binggeli \& Jerjen's 
intrinsic scatter interpretation. 

\section{Conclusions}

The findings presented in YC95 were based on the assumption that the 
relatively small scatters observed in two scaling relationships found for 
Fornax and Local Group dwarfs, are also applicable to Virgo dwarfs. Although, 
we still consider this assumption to be a most reasonable one, we concede that
we should have stated it explicitly in YC95. Unlike Binggeli \& Jerjen, we 
(YC95 \& this work) did not make any prior assumption as to the actual depth 
of Virgo when making our case. Also, contrary to the impression given by 
Binggeli \& Jerjen, our case is not undermined by any dependence between the 
scaling laws--which merely means that we slightly under-estimated our internal
errors in YC95. In this paper, we have also presented further statistical 
evidence to support our case based on Binggeli \& Jerjen's own dataset. 

Binggeli \& Jerjen claim to have presented evidence both that (1) profile 
shape is not a useful distance indicator because scaling laws based on it have 
large intrinsic scatters and that (2) the spatial depth of the Virgo Cluster 
must be small. However, all that they have in fact achieved in their 9-page
paper, is to point out that if one first assumes that the depth of the 
cluster is small enough, then there must be large intrinsic scatters in the 
scaling laws in question for Virgo galaxies. In order to explain away the 
smaller intrinsic scatters found in Fornax and Local Group samples of 
galaxies, they suggest [in the absence of any supporting evidence] that either
there is a dichotomy between Virgo and non-Virgo dwarfs or that our samples of 
non-Virgo dwarfs are not representative. They therefore cannot legitimately 
claim to have presented evidence for both (1) and (2) simultaneously; or to 
have presented evidence for (1) or (2) without first assuming (2) or (1) 
respectively.

The Virgo Cluster is not a suitable target for investigating the reliability
of the curvature-based indicators. Reliable photometry of galaxies in other 
clusters and groups, where there is no controversy concerning line-of-sight
depth, are required for this purpose.

\begin{acknowledgements}
CKY gratefully acknowledges a PDRF from the National Postdoctoral Fellowship 
Office of China and use of the computing facilities of the QSO \& 
Observational Cosmology Group at Beijing Astronomical Observatory.
\end{acknowledgements}

\end{document}